# Time Series Forecasting for Air Pollution in Seoul

Sean Jeon[a] and Seungmin Han[b]

[a]Department of Quantitative Theory and Methods, Emory University

[b]Johns Creek High School

**Abstract**

Accurate air pollution forecasting plays a crucial role in controlling air quality and minimizing adverse effects on human life. Among pollutants, atmospheric particulate matter (PM) is particularly significant, affecting both visibility and human health. In this study the concentration of air pollutants and comprehensive air quality index (CAI) data collected from 2015 to 2018 in Seoul, South Korea was analyzed. Using two different statistical models: error, trend, season (ETS) and autoregressive moving-average (ARIMA), measured monthly average $PM_{2.5}$ (particles with a diameter less than 2.5μm) concentration were used as input to forecast the monthly averaged concentration of $PM_{2.5}$ 12 months ahead. To evaluate the performance of the ETS model, five evaluation criteria were used: mean error (ME), root mean squared error (RMSE), mean absolute error (MAE), mean percentage error (MPE), and mean absolute percentage error (MAPE). Data collected from January 2019 to December 2019 were used for cross-validation check of ETS model. The best fitted ARIMA model was determined by examining the AICc (Akaike Information Criterion corrected) value. The results indicated that the ETS model outperforms the ARIMA model.

## 1. Introduction

Air pollution profoundly impacts human health, the economy, and environmental sustainability. It can lead to serious illnesses like lung cancer and heart disease. Monitoring air quality provides real-time data to control the cause of air pollution. Forecasting air quality, in addition, can provide information necessary for designing mitigation acts to prevent damage caused by air pollution (Zaini, 2022). Developing accurate forecasting models can improve air pollution forecasting and warnings, thus facilitating better control of pollution in Seoul. Therefore, the precise monitoring and forecasting is very important.

The main purpose of this study is to forecast air pollution in Seoul through the analysis of air pollutants, the comprehensive air-quality index (CAI), and meteorological variables. The CAI serves the purpose of making air quality easily understandable to the public and is calculated based on concentrations of $SO_2$, $CO$, $O_3$, $NO_2$, $PM_{10}$ and $PM_{2.5}$. In this study, statistic characteristics and correlations between six pollutants have been examined. Furthermore, meteorological variables which are highly related to six pollutants also have been explored. And then exponential smoothing and autoregressive integrated moving-average (ARIMA) methods have been applied to forecast the concentration of $PM_{2.5}$ in particular. For the analysis of the correlation between air pollutants and CAI, specific regions within Seoul – City-hall,

Ganseo-gu, Seocho-gu, and Sonpa-gu – have been selected. Ganseo-gu, among these regions, has been chosen for the development of forecasting models for $PM_{2.5}$. The real-time data measured every hour from 2015 to 2018 was used to develop forecasting models. For the evaluation of applied forecasting models, the data from 2019 to 2020 has been compared with in-sample data.

## 2. Relevant literature review

It has been demonstrated in previous studies that a range of variables and forecasting models are utilized for predicting air pollution. In this report, an exploration of the types of variables suitable for air pollution forecasting will be conducted through a literature review. Also, an examination will be undertaken of the various forecasting methods that have been utilized and developed in the existing literature.

### 2.1. Variables for forecasting

It has been observed from prior studies that a multitude of variables are employed for the prediction of air pollution. Air pollutants and meteorological variables commonly serve as independent factors for forecasting. For instance, PM10 and PM2.5 are significant air pollutants with the potential to significantly impact both human health and society. Elangasinghe (2014) conducted a time series analysis of $PM_{10}$ and $PM_{2.5}$ to forecast their concentrations in a coastal site. The application of an artificial neural network (ANN) model facilitated accurate forecasts of PM concentrations, which are heavily influenced by meteorological factors. Earlier studies have also been undertaken to enhance and develop existing forecasting models. Wang *et al*. (2015) developed a hybrid forecasting model for daily $PM_{10}$ and $SO_2$ concentrations. Given the close relationship between air pollutant concentrations and the geological and meteorological environment, correlated variables like wind speed can be simultaneously considered. Cogliani (2001) demonstrated the correlation between the air pollution index and meteorological variables such as thermic excursion, the previous day's air pollution index, and the daily average wind speed. Forecasting often involves using statistical metrics like mean pollutant concentrations. Kumar and Goyal (2011) utilized 24-hour average values of RSPM (Respirable Suspended Particulate Matter), $SO_2$, $NO_2$ and SPM (Suspended Particulate Matter). Moreover, indices such as the Air Quality Index (AQI), designed to portray air quality, are pivotal variables for air pollution forecasting. The AQI conveys air quality through categorized grades that differ among countries. Ganesh *et al*. (2017) employed regression models to predict AQI in Delhi, India, and Houston, America. The efficacy of air pollution forecasting has been demonstrated through the utilization of pollutant concentrations, statistical values for each variable, and various indices calculated using air pollutants.

### 2.2. Forecasting Models

The techniques and tools for real-time air quality forecasting can be categorized into three groups: simple empirical approaches, parametric or non-parametric approaches, and physically based approaches (Zhang *et al*., 2012). In this report, parametric or non-parametric approaches, rooted in statistical methods, will be examined through the literature review. The variables that exhibit a high correlation with air pollution can be individually examined (Anthes and Warner, 1978). Cogliani (2001) discovered that daily thermic excursion has a strong correlation with the air pollution index by analyzing measured meteorological variables. Statistical methods have been developed for air pollution forecasting and can be effectively used for predicting air quality. Gracia (2017) built four statistical models to forecast $PM_{10}$ concentration in the metropolitan area of northern Spain: Vector autoregressive moving-average (VARMA), autoregressive integrated moving-average (ARIMA), multilayer perceptron (MLP), neural networks and support vector machines (SNMs) with regression. Among these models, ARIMA, which explains autocorrelations in data, is one of the most widely utilized for air pollution forecasting. Numerous studies have been conducted to improve existing forecasting models. For instance, Chaloulakou, Saisana, and Spyrellis (2003) conducted a comparative assessment of forecasting models to predict summertime ozone levels in Athens. More recently, computational intelligence techniques like machine learning and deep learning technologies have been applied to air quality forecasting to enhance existing models (Zaini *et al.*, 2022).

## 3. Data analysis

### 3.1 Data

Concentrations of six types of air pollutants, along with CAI and meteorological variables, have been utilized for the analysis of air pollution in Seoul, South Korea. The major air pollutants include sulphur dioxide ($SO_2$), carbon monoxide (CO), ozone ($O_3$), nitrogen dioxide ($NO_2$) and particulate matter ($PM_{10}$ and $PM_{2.5}$). $SO_2$, a colorless gas, is emitted from volcanic eruptions and fossil fuel usage in industries and electricity generation (Makgato and Chirwa, 2020). CO, also emitted from fossil fuels, can have fatal effects on human health at high concentration levels (Chen et al., 2021). Ground-level $O_3$, formed from nitrogen oxide and volatile organic compounds, poses risks to both human health and the environment (Chen et al., 2021). $NO_2$, emitted from fossil fuels and power plants, is another significant air pollutant (Lu et al., 2021). $PM_{10}$ and $PM_{2.5}$ are fine atmospheric particles with diameters less than 10 μm and 2.5 μm, respectively. The CAI, calculated from the concentrations of the six pollutants mentioned above, provides a simplified representation of daily air quality. Additionally, meteorological variables are examined, which have strong correlation with pollutant concentrations.

The Ministry of Environment of the Republic of Korea provides real-time data on six air pollutants and the comprehensive air-quality index (CAI) from 614 monitoring points across 162 cities and counties. This study focuses on analyzing the concentrations of these six pollutants and CAI from four specific regions within Seoul (Table 1). The data has been collected every hour from each region. Monthly averaged values are used for forecast.

| Location | Address | Longitude | Latitude |
|---|---|---|---|
| City Hall | 15, Deoksugung-gil, Jung-gu, Seoul, Republic of Korea | 126.9747 | 37.5643 |
| Ganseo-gu | 71, Gangseo-ro 45da-gil, Gangseo-gu, Seoul, Republic of Korea | 126.8351 | 37.5447 |
| Banpo-dong | 16, Sinbanpo-ro 15-gil, Seocho-gu, Seoul, Republic of Korea | 126.9945 | 37.5046 |
| Bang-yi-dong | 59, Gucheonmyeon-ro 42-gil, Gangdong-gu, Seoul, Republic of Korea | 127.1368 | 37.545 |

**Table 1. Selected regions for analyzing and forecasting**

### 3.2 Statistical characteristics of variables

Strong seasonal trends in the concentrations of six air pollutants are observed at four distinct measuring points. The average values of these six air pollutants and CAI are presented in Table 2. The data reveals that Ganseo-gu exhibits the highest concentrations in both CAI and the majority of air pollutants among the four regions.

|  | SO$_2$(ppm) | CO(ppm) | O$_3$(ppm) | NO$_2$(ppm) | PM$_{10}$(μg/m$^3$) | PM$_{2.5}$(μg/m$^3$) | CAI(Index) |
|---|---|---|---|---|---|---|---|
| City-hall | 0.004 | 0.531 | 0.024 | 0.034 | 41.150 | 23.384 | 84.031 |
| Ganseo_gu | 0.005 | 0.465 | 0.025 | 0.030 | 46.264 | 24.424 | 85.995 |
| Seocho_gu | 0.004 | 0.469 | 0.024 | 0.029 | 46.102 | 23.622 | 84.737 |
| Songpa_gu | 0.004 | 0.511 | 0.021 | 0.031 | 46.178 | 24.350 | 85.339 |

Table 2. Average of air pollutants and CAI concentrations

The following figures show the concentrations of air pollutants in each month. The graph indicates that SO$_2$ concentration is elevated during the spring period, especially in February and March. In Ganseo-gu, however the SO$_2$ is exceptionally higher in April than in February.

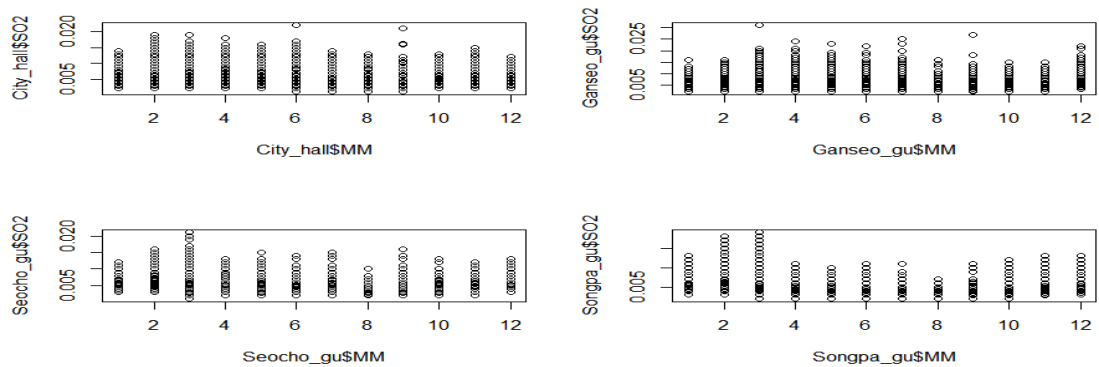

Figure 1. Seasonality of SO$_2$

The trend of CO concentration is similar to that of SO2 in that it is high when the temperature is low (e.g. winter season) as in Figure 2. The CO, however, exhibits clearer trend than SO$_2$ in general displaying lower concentration in summer season.

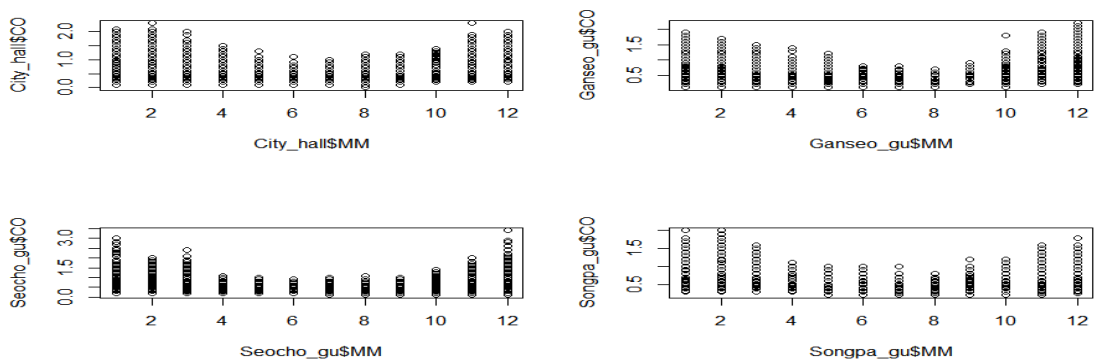

Figure 2. Seasonality of CO

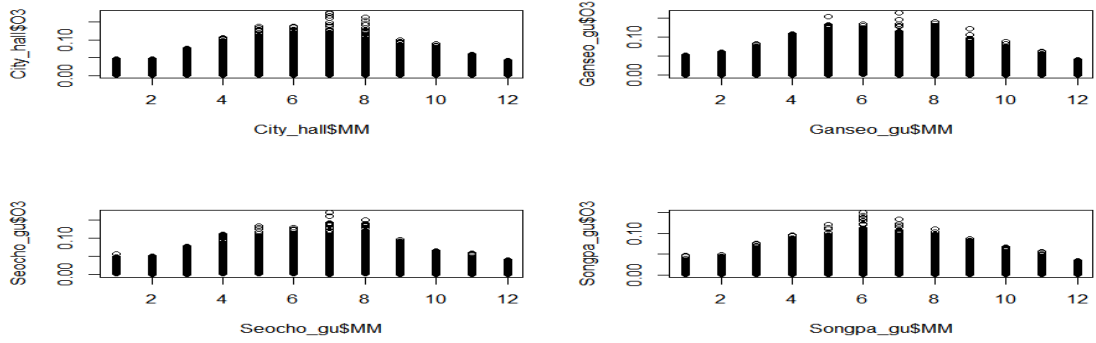

**Figure 3. Seasonality of $O_3$**

Figure 3 represents that the concentration of $O_3$ is high during the summer. The overall trends are opposite to those of $SO_2$ and CO. In Figure 4, $NO_2$ concentration shows the weakest trend among the air pollutants, while the $NO_2$ concentration is lower when the temperature is high.

Similar patterns are observed in Figures 5 and 6 for $PM_{10}$ and $PM_{2.5}$, wherein the concentration is high in spring. However, $PM_{2.5}$ maintains high concentration throughout the year, while $PM_{10}$ exhibits a significantly higher trend in February.

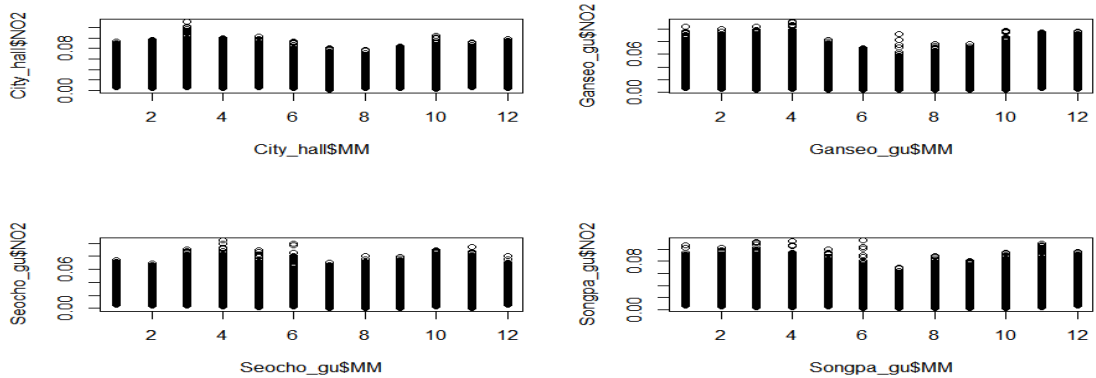

**Figure 4. Seasonality of $NO_2$**

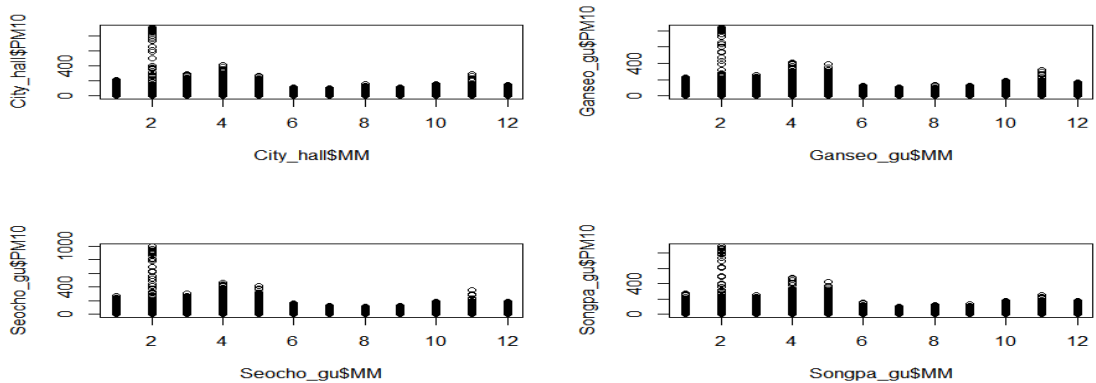

**Figure 5. Seasonality of $PM_{10}$**

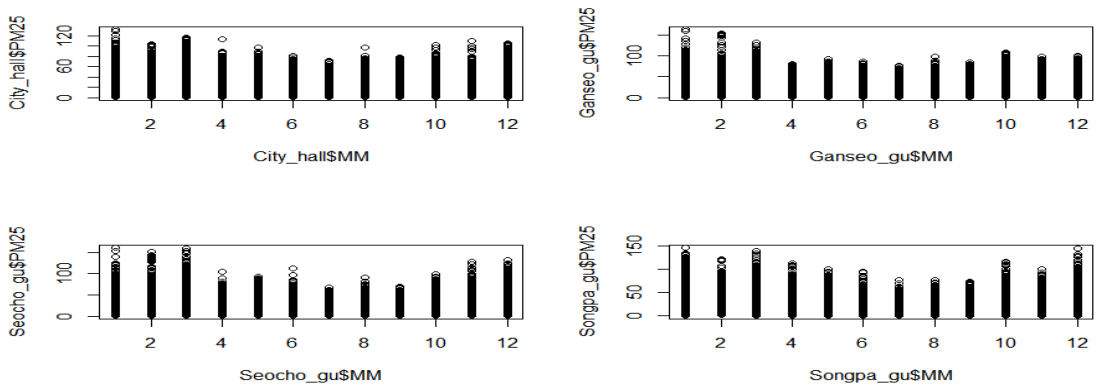

**Figure 6. Seasonality of PM$_{2.5}$**

*3.3 Correlation between variables*

Among the six air pollutants, it is acknowledged that PM$_{2.5}$ not only poses risks to human health but also significantly affects the quality of life due to its association with visibility (Cobourn, 2010; McKeen et al., 2007). In this report, the focus will be on examining the correlation between PM$_{2.5}$ and other variables for the purpose of air pollution forecasting. As demonstrated in the graph above, the PM$_{2.5}$ concentration is elevated during the cold and windy spring season. Table 3 presents the correlation of PM$_{2.5}$ with the other five air pollutants. Among these pollutants, PM$_{10}$ exhibits the highest correlation with PM$_{2.5}$ in each selected region. The data indicates that CO possesses the next highest correlation, while SO$_2$ and NO$_2$ also maintain relatively strong correlations compared to O$_3$. Among the meteorological variables, temperature demonstrates a notably high correlation with PM$_{2.5}$ (Table 4). Wind speed and precipitation also exhibit comparatively high correlations with PM$_{2.5}$.

|  | SO$_2$ (ppm) | CO (ppm) | O$_3$ (ppm) | NO$_2$ (ppm) | PM$_{10}$ (µg/m$^3$) |
|---|---|---|---|---|---|
| **City hall** | 0.405 | 0.553 | 0.015 | 0.450 | 0.694 |
| **Ganseo_gu** | 0.384 | 0.555 | -0.013 | 0.429 | 0.729 |
| **Seocho_gu** | 0.406 | 0.557 | -0.001 | 0.424 | 0.724 |
| **Songpa_gu** | 0.364 | 0.678 | -0.082 | 0.527 | 0.702 |

**Table 3. Correlation between PM$_{2.5}$ and other air pollutants**

| Location | Temperature(°C) | Precipitation(mm) | Wind speed(m/s) | Local air pressure(hPa) | Visibility(10m) |
|---|---|---|---|---|---|
| **City hall** | -0.095 | -0.061 | -0.096 | 0.062 | -0.592 |
| **Ganseo-gu** | -0.094 | -0.073 | -0.132 | 0.058 | -0.557 |
| **Seocho-gu** | -0.136 | -0.068 | -0.092 | 0.073 | -0.561 |
| **Songpa-gu** | -0.188 | -0.070 | -0.122 | 0.101 | -0.515 |
| **Average** | -0.128 | -0.068 | -0.110 | 0.073 | -0.556 |

**Table 4. Correlation between PM$_{2.5}$ and other air pollutants**

*3.4 Trend and seasonality of PM$_{2.5}$*

Ganseo-gu, which shows the highest PM$_{2.5}$ concentration, has been selected to further analysis and forecast. Figure 7 illustrates the trend, seasonal and random components of PM$_{2.5}$ in Ganseo-gu. distinctly depicts a declining tendency from the third quarter of 2015 to the third quarter of 2016, followed by an increasing pattern for the rest of the period. In terms of seasonality, it becomes evident that the concentration is elevated during the spring and winter seasons, characterized by relatively cold weather and low air temperatures.

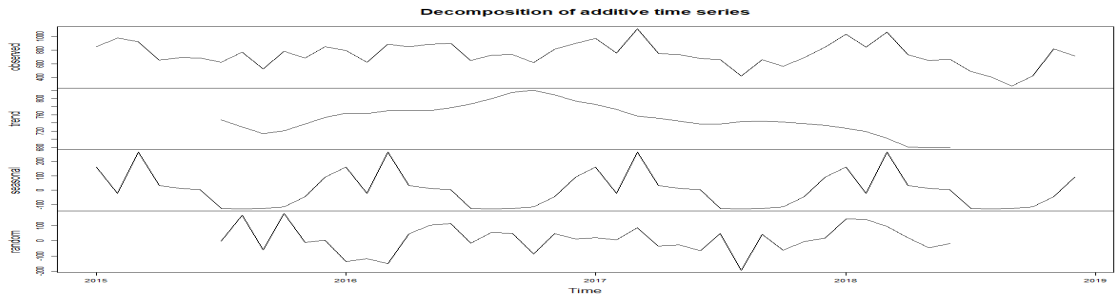

Figure 7. Trend, seasonal and random components of PM$_{2.5}$ concentration

## 4 Forecasting models

In this study, exponential smoothing and autoregressive integrated moving-average (ARIMA) models have been applied to forecast PM$_{2.5}$ concentration in Ganseo-gu. In this report, formulas and some application cases of exponential smoothing and ARIMA models will be examined.

### 4.1. Error, trend, seasonal (ETS) model

The exponential smoothing method was suggested in late 1950 (Brown, 1959; Holt, 1957; Winters, 1960). It is based on the idea that the average of past observations affects future values with higher weights on more recent values (Hyndman and Athanasopoulos, 2018). In other words, the weight assigned to observations diminishes exponentially as they become older. The ETS model, which is a statistical model underlying exponential smoothing, was applied to forecast PM$_{2.5}$ in this paper. The ETS model accounts for variations using both additive and multiplicative approaches within the trend and seasonal components (Pegels, 1969). The parameters used for ETS model are listed in table 5 (Hyndman and Athanasopoulos, 2018)

| Components | Methods |
| --- | --- |
| **Error** | A(additive), M(multiplicative) |
| **Trend** | N(none), A(additive), Ad(additive damped), M(multiplicative), Md(multiplicative damped) |
| **Seasonal** | N(none), A(additive), M(multiplicative) |

Table 5. Parameters used for error, trend, seasonal (ETS) model

Three forecasting models were employed by Cekim (2020) for the prediction of PM10 concentration across 18 cities in Turkey: error, trend, and seasonal (ETS); autoregressive integrated moving average (ARIMA); and singular spectrum analysis (SSA). To assess the appropriateness of each forecasting model in every city, the Root Mean Square Error (RMSE) was utilized. In each city, the most effective model was implemented, and the study proposed forecasted values.

### 4.2 Autoregressive integrated moving-average (ARIMA) model

Autoregressive integrated moving-average (ARIMA) model is one of the most commonly used approaches to time series forecasting (Hyndman and Athanasopoulos, 2018). ARIMA is composed of autoregressive (AR) model and moving average (MA) model with the degree of differences to achieve stationary. The autoregressive model of order p can be denoted as (Hyndman and Athanasopoulos, 2018):

$$y_t = c + \varphi_1 y_{t-1} + \varphi_2 y_{t-2} + \cdots + \varphi_p y_{t-p} + e_t \quad (1)$$

The moving average model of order q can be represented as (Hyndman and Athanasopoulos, 2018):

$$y_t = c + e_t + \theta_1 e_{t-1} + \theta_2 e_{t-2} + \cdots + \theta_q e_{t-q} \quad (2)$$

As a result, the autoregressive moving average (ARMA) model can be expressed as (Hyndman and Athanasopoulos, 2018):

$$y'_t = c + \varphi_1 y'_{t-1} + \varphi_2 y'_{t-2} + \cdots + \varphi_p y'_{t-p} + e_t + \theta_1 e_{t-1} + \theta_2 e_{t-2} + \cdots + \theta_q e_{t-q} \quad (3)$$

when $y'_t$ is the differenced series. The equation of ARIMA model including differences can be expressed as (Hyndman and Athanasopoulos, 2018):

$$(1 - \varphi_1 B - \cdots - \varphi_p B^p)(1 - B)^d y_t = c + (1 + \theta_1 B + \cdots + \theta_q B^q) e_t \quad (4)$$

Gracia (2017) forecasted $PM_{10}$ concentration in northern Spain using four different models: autoregressive integrated moving-average (ARIMA), vector autoregressive moving-average (VARMA), multilayer perceptron (MLP) and neural networks and support vector machines (SVMs) with regression. It the study, ARIMA model demonstrated effective performance in both short term (one month) and comparatively long term (several months) period of forecasting.

## 5   Results and Discussion

*5.1. Results of Error, trend, seasonal (ETS) model*

In this study, the application of five ETS models to predict the concentration of PM2.5 in Ganseo-gu, Seoul, has been carried out: ANN, AAN, AAN damped, AAA, and AAA damped. Figure 8 shows the result of forecast using each model.

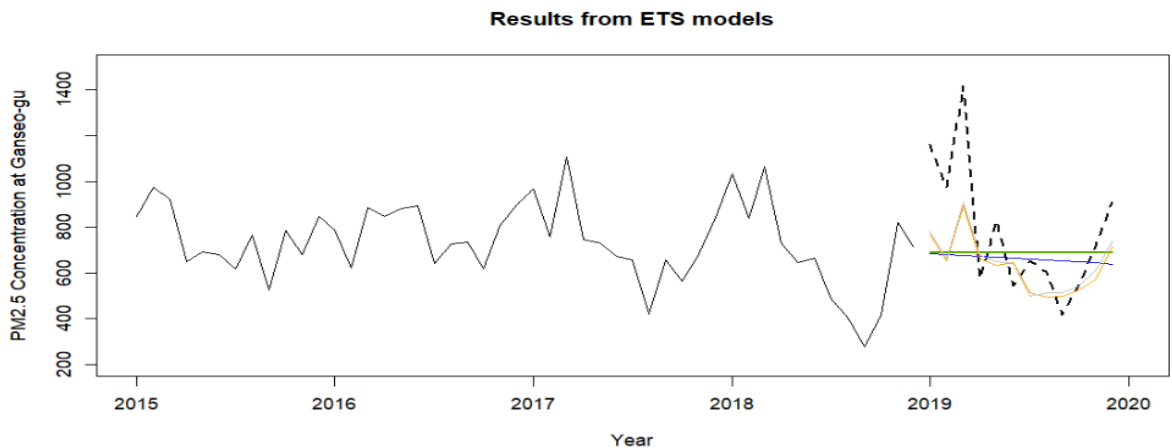

**Figure 8. Results from error, trend, seasonal (ETS) models**

To assess the performance of ETS models, evaluation criteria have been considered: mean error (ME), root mean squared error (RMSE), mean absolute error (MAE), mean percentage error (MPE), mean absolute percentage error (MAPE). Among the applied models, AAA damped model demonstrates the best fit for both training and test sets, based on all of the criteria.

| Models | Data set | ME | RMSE | MAE | MPE | MAPE |
| --- | --- | --- | --- | --- | --- | --- |
| ANN | Training set | -6.159 | 161.863 | 133.532 | -4.825 | 19.821 |
|  | Test set | 86.922 | 293.186 | 221.636 | 0.367 | 26.971 |
| AAN | Training set | -0.525 | 161.776 | 133.545 | -3.985 | 19.686 |
|  | Test set | 116.452 | 296.549 | 218.268 | 4.926 | 25.427 |
| AAN damped | Training set | -3.520 | 161.606 | 133.457 | -4.469 | 19.764 |
|  | Test set | 88.679 | 293.586 | 221.563 | 0.626 | 26.894 |
| AAA | Training set | -2.329 | 112.831 | 93.151 | -3.855 | 14.608 |
|  | Test set | 147.288 | 238.507 | 192.212 | 13.232 | 22.076 |
| AAA damped | Training set | -11.400 | 112.850 | 92.564 | -5.089 | 14.749 |
|  | Test set | 133.950 | 229.923 | 182.985 | 11.349 | 21.096 |

Table 5. Evaluation criteria values of error, trend, seasonal (ETS) models

For the further verification, the cross-validation check has been conducted. The data was divided into in-sample period including from 2015 to 2018 and out of sample including from 2019. For the five ETS models referred above, mean absolute percentage error (MAPE) were 26.97108, 25.42727, 26.89370, 34.72839, and 24.03151, respectively. The outcome of the cross-validation process further verifies that the AAA damped model demonstrates the most superior performance.

*5.2. Results of autoregressive integrated moving-average (ARIMA) model*

The autoregressive intergrated moving-average (ARIMA) model assumes that the data is stationary in the mean and variance. Figures 6 and 7 display the concentration of $PM_{2.5}$ have seasonality and trend. The ACF and PACF plots in Figure 9 also display the time series of $PM_{2.5}$ concentration is not stationary, showing much higher values than significant value at lag 1, 2, 5, 6, 7, and 12. To make the time series stationary, seasonal differencing was implemented. The resulting ACF and PACF plots, as illustrated in Figure 10, demonstrate a rapid decline to zero in both ACF and PACF.

With the differenced data, ARIMA models with five different parameters have been applied: where (p,d,q) is (2,0,0), (3,0,0), (2,1,0), (2,2,0), (2,1,1), and (2,1,2). Based on AICc values, it is represented that ARIMA(2,1,1) model is best fitted, showing AICc values of 589.26, 591.55, 583.66, 593.72, 580.73, 580.73, and 582.55 respectively. By evaluating AICc values, it becomes evident that the ARIMA(2,1,1) model exhibits the best fit, showcasing AICc values of 589.26, 591.55, 583.66, 593.72, 580.73, 580.73, and 582.55, respectively. The graph depicted in Figure 11 displays the forecasted values for the next 12 months as generated by the ARIMA(2,1,1) model. Figure 12 displays the outcomes of ARIMA(2,1,1), demonstrating that the residuals exhibit characteristics of white noise, indicating an absence of autocorrelation within this model.

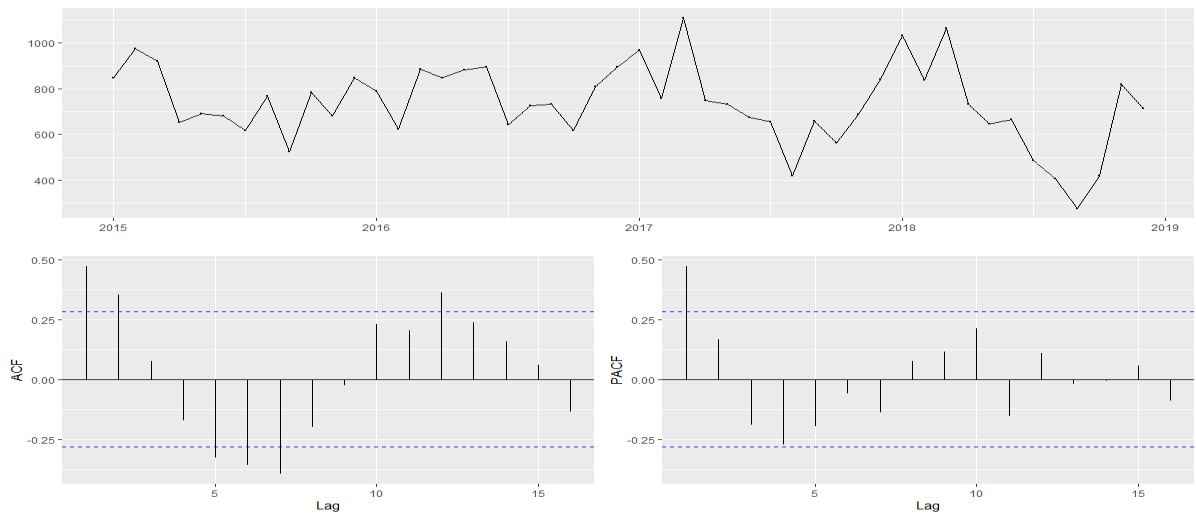

**Figure 9. ACF and PACF plots of PM$_{2.5}$ concentration data**

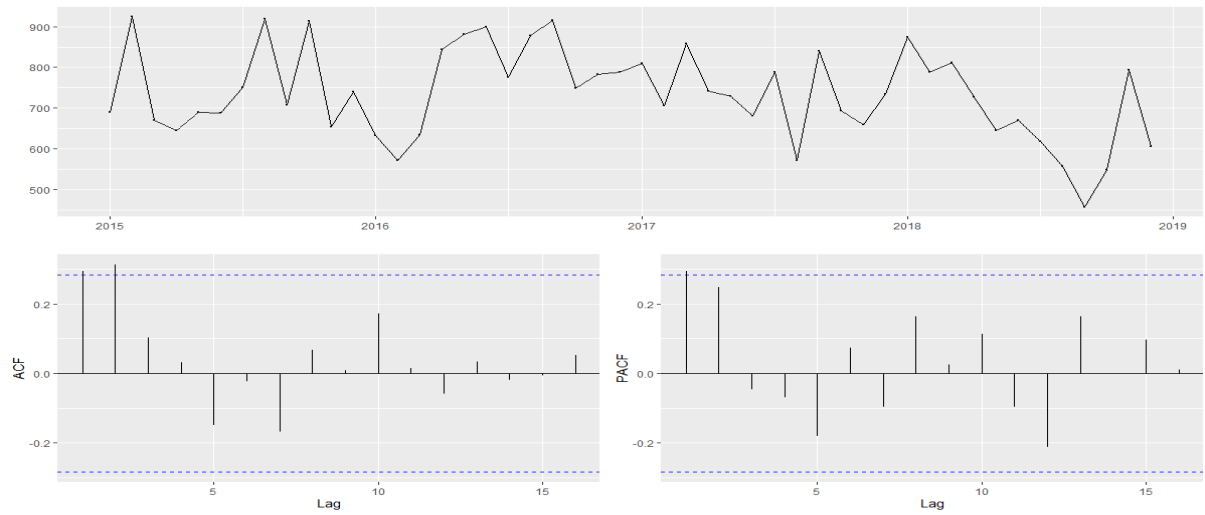

**Figure 10. ACF and PACF plots of differenced PM$_{2.5}$ concentration data**

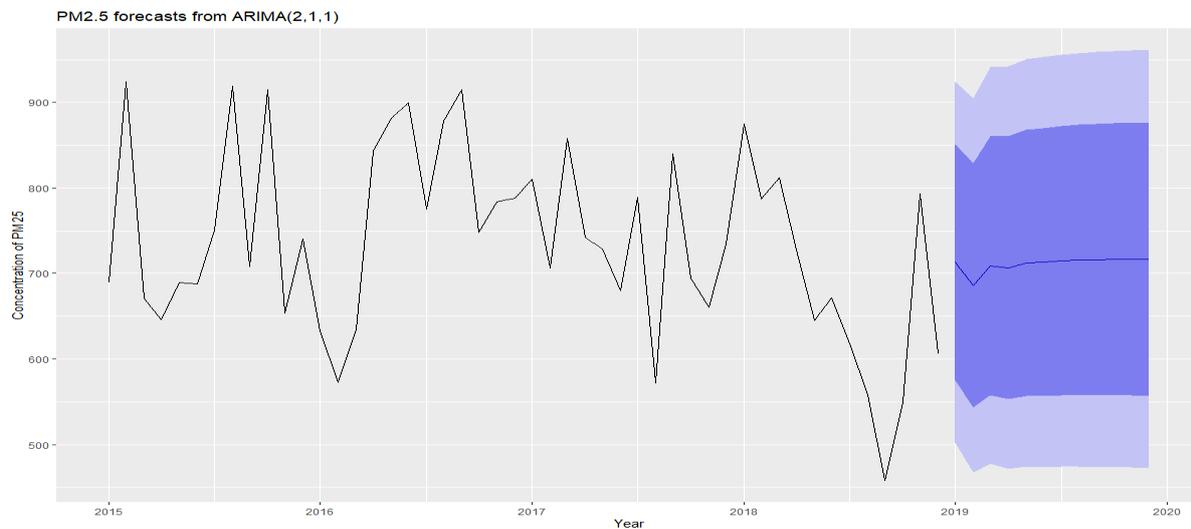

**Figure 11. Results from the ARIMA(2,1,1) model**

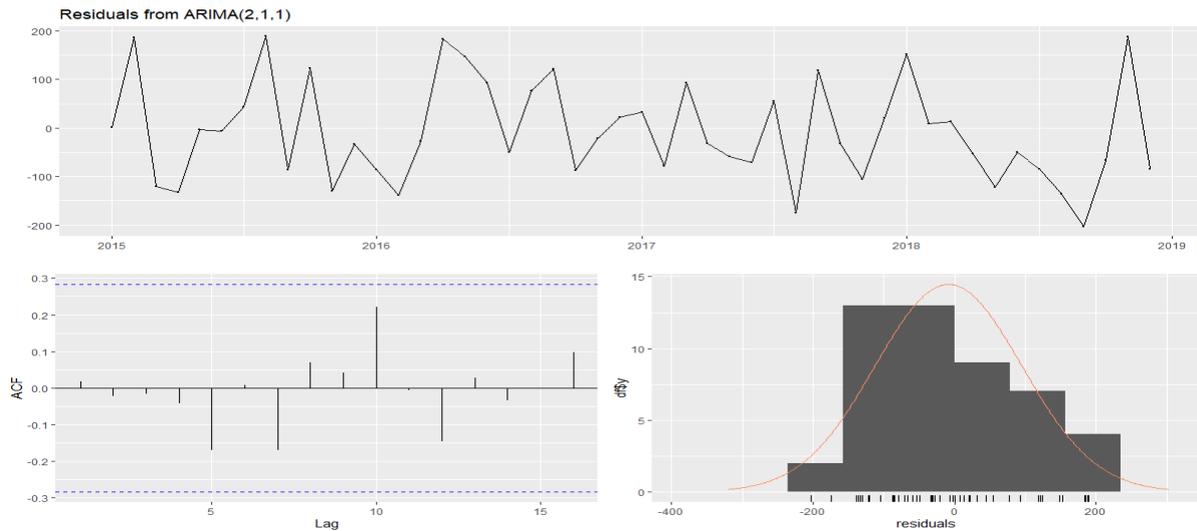

**Figure 12. Residuals of the ARIMA(2,1,1) model**

### 5.3. Comparison of ETS and ARIMA models

Some of the evaluation criteria, such as MAPE and AICc, have been employed to facilitate the comparison of forecasting models' performance. AICc becomes particularly helpful when dealing with models of the same category. For instance, when assessing various ARIMA candidate models featuring distinct parameters, AICc serves as a beneficial evaluation criterion. However, AICc lacks applicability when contrasting the performance of diverse forecasting models. Furthermore, the comparison of evaluation criteria for ETS and ARIMA remains valid only when employing data of the same orders of differencing.

In this study, therefore, a comparison was made between the ETS(A,Ad,A) model and the ARIMA(0,0,0)(0,1,2) model, with both models employing undifferenced data. Figure 13 and Figure 14 present the outcomes of the ETS(A,Ad,A) and ARIMA(0,0,0)(0,1,0) models, respectively. The evaluation criteria-based analysis indicates that the ETS(A,Ad,A) model outperforms the ARIMA model in forecasting PM2.5 concentration (Table 6).

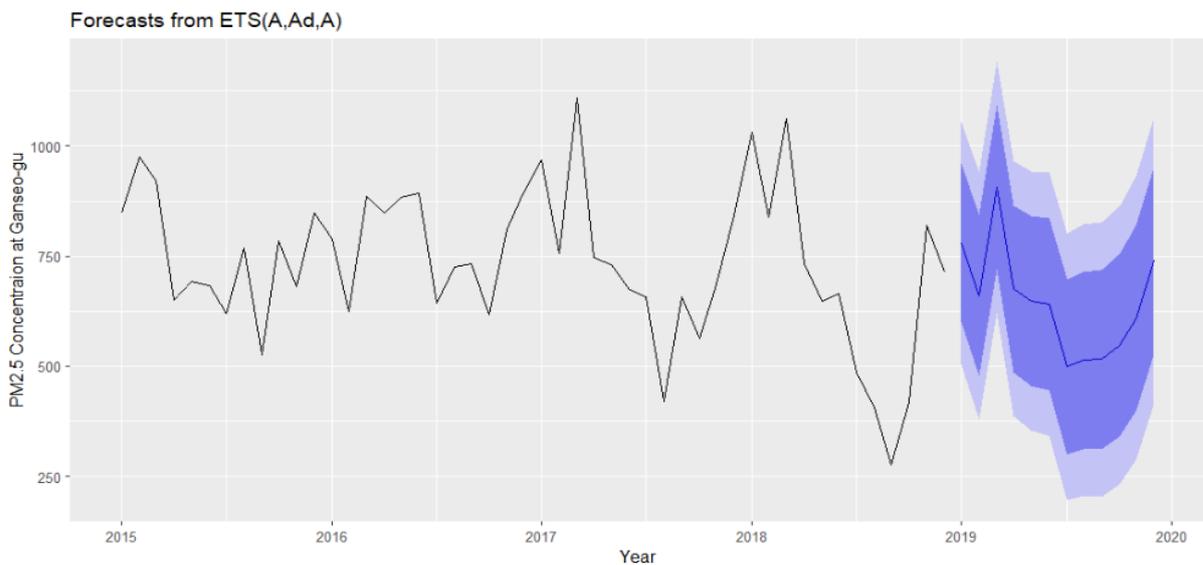

**Figure 13. Results from ETS(A,Ad,A) model**

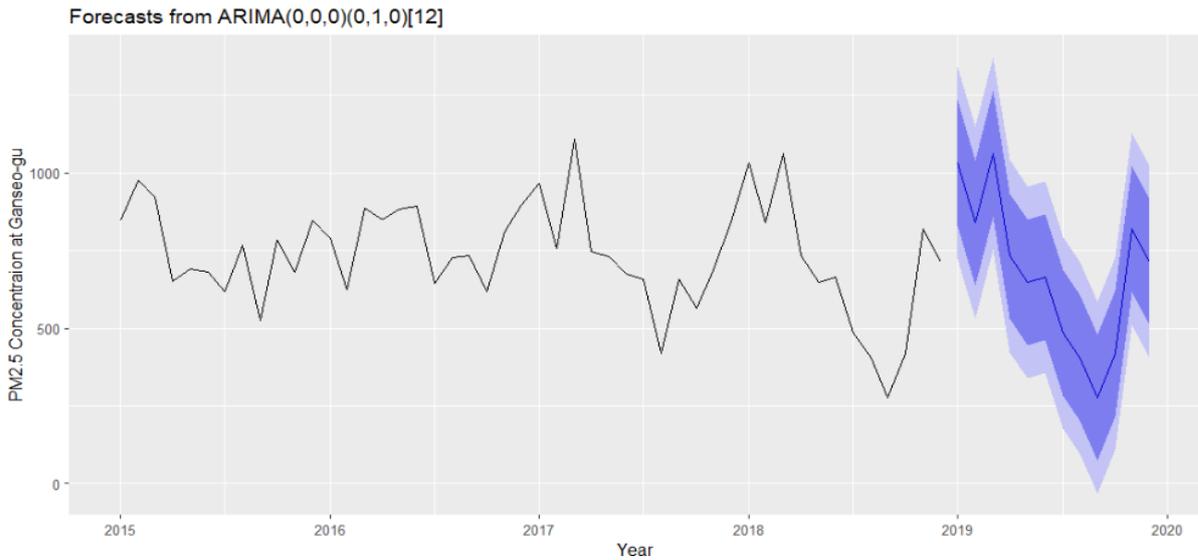

Figure 14. Results from ARIMA(0,0,0)(0,1,0) model

| Models | Dataset | ME | RMSE | MAE | MPE | MAPE |
|---|---|---|---|---|---|---|
| ETS(A,Ad,A) | Training set | -0.605 | 4.608 | 3.190 | -6.768 | 15.791 |
|  | Test set | 3.358 | 5.837 | 5.489 | 12.043 | 22.768 |
| ARIMA(0,0,0)(0,1,0) | Training set | 0.111 | 3.651 | 2.996 | -2.698 | 14.028 |
|  | Test set | 5.073 | 7.732 | 6.148 | 14.915 | 21.189 |

Table 6. Evaluation criteria values of ETS(A,Ad,A) and ARIMA(0,0,0)(0,1,0) model

## 6 Conclusions

In this report, the monthly average concentration of $PM_{2.5}$ in Ganseo-gu, Seoul, South Korea, has been forecasted for a 12-month period (from January 2019 to December 2019) using two distinct models: error, trend, seasonal (ETS) and autoregressive moving-average (ARIMA) model. The results indicate that both models are highly effective in predicting the monthly $PM_{2.5}$ concentration. Nonetheless, a comparison of the two models reveals that the ETS model (RMSE = 5.837 for the test set) outperforms the ARIMA model (RMSE = 7.732 for the test set).

In summary, considering its superior performance, the ETS model can be extended to other regions like City-hall, Seocho-gu, and Songpa-gu. For future studies, variables highly correlated with $PM_{2.5}$, such as wind speed and temperature, could be considered to enhance forecasting accuracy.